\documentclass[useAMS,usenatbib]{mn2e}

\usepackage{graphicx}
\usepackage{natbib}

\title[The velocity dispersion profile of $\omega$ Cen]{The non-peculiar velocity dispersion profile of the stellar system $\omega$
Centauri\thanks{Based on FORS observations collected with the Very Large Telescope
at the European Southern Observatory, Cerro Paranal, Chile, within the
observing programs 071.D-0217A and 081.D-0255A.}}
\author[A. Sollima et al.]{A. Sollima$^{1}$\thanks{E-mail:
asollima@iac.es}, M. Bellazzini$^{2}$, R. L. Smart$^{3}$, M. Correnti$^{4}$, E.
Pancino$^{2}$,
\newauthor
F. R. Ferraro$^{4}$ and D. Romano$^{4}$\\
$^{1}$ Instituto de Astrofisica de Canarias, c/via Lactea s/n, San Cristobal de
La Laguna, 38205, Spain\\
$^{2}$ INAF Osservatorio Astronomico di Bologna, via Ranzani 1, Bologna, 40127,
Italy\\
$^{3}$INAF, Osservatorio Astronomico di Torino, via Osservatorio 20, 10025 Pino
Torinese, Italy\\
$^{4}$Dipertimento di Astronomia, Universit\`a di Bologna, via Ranzani 1, Bologna, 40127,
Italy}
\begin{document}

\date{Accepted 2009 February 11. Received 2009 February 11; in original form
2009 February 11}

\pagerange{\pageref{firstpage}--\pageref{lastpage}} \pubyear{2009}

\maketitle

\label{firstpage}

\begin{abstract}
We present the results of a survey of radial velocities over a wide region
extending from $r\simeq 10\arcmin$ out to $r\simeq 80\arcmin$ ($\sim$1.5 tidal
radii) within the massive star cluster $\omega$ Centauri. The survey was
performed with FLAMES@VLT, to study the velocity dispersion
profile in the outer regions of this stellar system.
We derived accurate radial velocities for a sample of 2557 newly
observed stars, identifying 318 bona-fide cluster red giants. Merging our
data with those provided by Pancino et al. (2007), we assembled a final
homogeneous sample of 946 cluster members that allowed us to trace the
velocity dispersion profile from the center out to $r\sim$32 arcmin. The
velocity dispersion appears to decrease monotonically over this range, from a
central value of $\sigma_{v}\sim~17.2~Km~s^{-1}$ down to a minimum value of
$\sigma_{v}\sim~5.2~Km~s^{-1}$. The observed surface brightness profile,
rotation curve, velocity dispersion profile and ellipticity profile are
simultaneously well reproduced by a simple dynamical model in which mass
follows light, within the classical Newtonian theory of gravitation. The
comparison with an N-body model of the evolution of a system mimicking 
$\omega$ Cen during the last 10 orbits into the Galactic potential suggests
that (a) the rotation of stars lying in the inner $\simeq 20\arcmin$ of the
clusters is not due to the effects of the tidal field of the Milky Way, as
hypothized by other authors, and (b) the overall observational 
scenario is still compatible with the possibility that the outer regions of the
cluster are subject to some tidal stirring.

\end{abstract}

\section{Introduction}
\label{intro}

$\omega$ Centauri is the most massive and luminous globular cluster (GC) of the
Milky Way ($M \sim 2.5 ~ 10^6~M_{\odot}$; van de Ven et al. 2006). It is the
only known Galactic GC which shows clear star-to-star variations in the
abundance of iron-peak elements (Freeman \& Rodgers 1975; Norris, Freeman \&
Mighell 1996) and indication exists for possible helium abundance variation 
among its stellar populations (Norris 2004; Piotto et al. 2005). 
Together with a few other clusters (like M54 and NGC2419, for example),
$\omega$ Cen fall well above the sharp upper envelope of the main distribution
of GCs on the size-luminosity plane (Mackey \& van den Bergh 2005; Federici et
al. 2007). These (and other) evidences suggest that $\omega$ Cen could be not a
"genuine" GC but more likely the nuclear remnant of a dwarf galaxy that merged
in the past with the Milky Way (Freeman 1993; see Bellazzini et al. 2008 for a
discussion of the analogy with the case of M54 within the disrupting 
Sgr dwarf galaxy). 

From the dynamical point of view, $\omega$ Cen is one of the most flattened
among Galactic GCs (Meylan 1987). The flattening is generally interpreted as
due to the observed rotation, as the variations of ellipticity with distance
from the cluster center correlates with the amplitude of the rotation curve
(Merritt, Meylan, \& Mayor 1997; Meylan \& Mayor 1986). Merritt et al.(1997)
found that $\omega$ Cen can be well described as an isotropic oblate rotator
in which the mass is distributed as the light. In a recent study Van de Ven et
al. (2006) modelled $\omega$ Cen with an axisymmetric implementation of
Schwarzschild's orbit superposition method. They found that the system is
close to isotropic inside a radius of about 10 arcmin and becomes increasingly
tangentially anisotropic in the outer region, which displays significant mean
rotation. These authors suggested that this phase-space structure could be
caused by the effects of the tidal field of the Milky Way, as kinematical
gradients are a typical outcome of tidal stresses (see Mu{\~n}oz et al. 2008 and
references therein).

A number of studies on the kinematics of the cluster have been performed in
the recent past (Meylan et al. 1995; Mayor et al. 1997; Norris et al. 1997;
Reijns et al. 2006; Van de Ven et al. 2006), particularly focused on the
innermost $r\la 20\arcmin$ region. Recently, Scarpa, Marconi \& Gilmozzi
(2003) measured the velocity dispersion profile of $\omega$ Cen at larger
distances ($20'<r<28'$) using UVES spectra of 75 candidate cluster members. 
They found that the velocity dispersion profile remains constant at large radii
rather than decrease monotonically, as expected for systems in which mass
follows light. These authors suggested that this behaviour could be due to a
breakdown of Newton's law in the weak acceleration regime, as proposed in the
popular Modified Newtonian Dynamics framework (MOND, Milgrom 1983, 2008). 
However, it has been noted that actually this cannot be the case, as a flat
velocity profile at large radii is predicted by MOND only for clusters in the
deepest MOND regime, i.e. where both the internal accelleration and that due to
the gravitational field of the Milky Way are significantly lower that the MOND
scale accelleration, $a_0\simeq 1.2\times 10^{-8}$ cm/s$^2$ (Baumgardt, Grebel
\& Kroupa 2005; Moffat \& Toth 2008). 
On the contrary, at the position of $\omega$ Cen the external
accelleration due to the Milky Way is $\ga a_0$. In this regime, the
predictions of MOND are similar to the Newtonian behaviour (see also Moffat \&
Toth 2008; Haghi et al. 2009). 

Nevertheless, a flat velocity dispersion profile over a large radial range
would be uncompatible with a mass-follow-light model (Gilmore et al. 2007) and
may suggest the presence of a halo of Dark matter (DM) embedding the star
cluster and driving the kinematics of the stars in the outer regions (Carraro
\& Lia 2000; Mashchenko \& Sills 2005). In any case the available evidences are
not sufficient to support the case of DM: as McLaughlin \& Meylan (2003)
demonstrated, Scarpa et al.'s data could be well described by simple,
self-consistent dynamical models without the need of MOND and/or dark matter.
A larger kinematic survey in the region $r>20\arcmin$ is overdue to assess the
behaviour of the dispersion profile in this range. 

A controversial issue is also represented by the possible presence of signatures of
tidal effects in the outskirts of this stellar system. If the present-day
cluster is just the nuclear remnant of a larger system, it is reasonable to
expect some observable overdensity of stripped stars in the surroundings, if the
latest phase of disruption occurred in the recent past (up to a 
few orbital period ago; see Combes, Leon \& Meylan 1999; Dinescu (2002);
Tsuchiya, Korchagin \& Dinescu 2004; 
Mizutani, Chiba \& Sakamoto 2003; Ideta \& Makino 2004; Bekki \& Freeman 2003).
Leon, Meylan \& Combes (2000) studied the 2-D structures of the distribution of stars 
around $\omega$ Cen from $5.5^{o}\times 5.5^{o}$ photographic films obtained 
with the ESO Schmidt telescope. 
This analysis evidenced the presence of a pair of tidal tails approximately
oriented in the direction of the Galactic center, in projection. This result
has been questioned by Law et al. (2003) who found that {\it a)} Leon's et al.
tidal tails were strongly correlated with unhomogeneities in the reddening
distribution and {\it b)} no extra-tidal component can be detected using the
homogeneous near-infrared photometry of the 2MASS survey. 
In a recent dedicated survey, Da Costa \& Coleman (2008) derived the radial
velocities (with an accuracy of $\sim 10~Km~s^{-1}$) of 4105 stars located on a wide
region out to a distance of 2 tidal radii from the cluster center. They
estimated that less than 0.7\% of the total cluster mass is comprised
between 1 and 2 tidal radii, implying that the stripping process on the
progenitor (if any) must have been largely completed at early epochs.

In this paper we present the results of a survey of radial velocities in
$\omega$ Cen performed with FLAMES@VLT, focused on the radial range $10\arcmin
\la r\la 30\arcmin$ and aimed at the assessment of the kinematical properties
in the outskirts of the cluster. The new dataset, combined with the
homogeneopus sample of Pancino et al. (2007), allowed us to derive a robust
velocity dispersion profile from the center to a distance of $32\arcmin$ with an
accuracy $\leq 1~Km~s^{-1}$. In \S 2 we describe the observations and the data
reduction techniques. In \S 3 we present the metallicity and the velocity
distributions. Section 4 is devoted to the description of the method used to
derive the velocity dispersion profile of $\omega$ Cen. In \S 5 we compare the
observed kinematics of this stellar system with a set of dynamical models and
N-body simulations. Section 6 is devoted to the analysis of the sample of stars
located in the outer region of $\omega$ Cen. Finally, we summarize and discuss
our results in \S 7.

\section{Observations and Data reduction}

\begin{figure}
 \includegraphics[width=8.7cm]{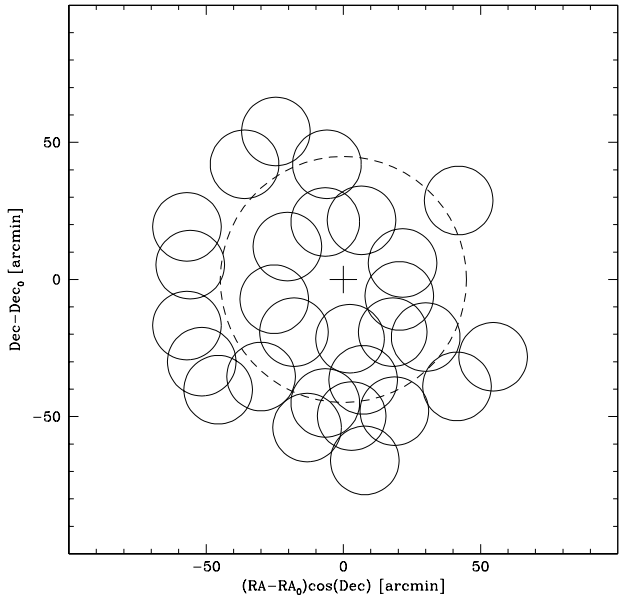}
\caption{Map of the region sampled by the FLAMES observations. North is up,
east towards the right-hand side. The 28 fields observed with FLAMES ({\it
outer sample}) are shown as continuous circles. The cluster center and tidal
radius (from Trager et al. 1995) are indicated by the black cross and the
dashed line, respectively.}
\label{map}
\end{figure}

The analysis presented here is based on two spectroscopic datasets: {\it i)} the
{\it inner sample}: constituted of a sample of $\sim$ 700 red giants selected
from the photometry by Pancino et al. (2000) in the magnitude range $13<B<16$
lying within $15\arcmin$ from the cluster center and presented in
Pancino et al. (2007; hereafter P07), and {\it ii)} 
the {\it outer sample}: constituted of a sample of 2557 red giants selected 
from the 2MASS catalog (Skrutskie et al. 2006) between $6.0<K<13.0$, observed in 
28 pointings at distances $10'<r<80'$ from the cluster center 
(see Figure \ref{map}). 

At $r\sim 20\arcmin$ from the cluster center, the surface brightness is already
$\mu_V>22.0$ mag/arcsec$^2$, implying that Red Giant Branch (RGB) stars of
$\omega$ Cen should be relatively rare and indeed not numerous enough to fill
all the fibers of a single FLAMES plate. To achieve the maximum efficiency in
selecting cluster members for the {\em outer sample} 
we gave the highest priority for fiber allocation to
stars (a) lying within a selection box in color-magnitude diagram broadly
enclosing the cluster RGB (see Fig.~\ref{cmd}), defined from the accurate
infrared photometry by Sollima et al. (2004), and (b) with UCAC2 (Zacharias et
al. 2004) proper motions differing by less than $\pm~8~mas~yr^{-1}$ from the
bulk proper motion of the cluster, as determined by van Leeuwen et al. (2000).
These "a priori" criteria allowed to select some hundreds high-confidence
target stars. The remaining fibers were positioned on other stars selected
from the 2MASS catalog in the same magnitude range, as shown in Fig.~\ref{cmd},
below. 

Observations have been done with FLAMES (Pasquini et al. 2002) at the ESO VLT
in Paranal, Chile, between 2003 May 22 and 28 ({\it inner sample}, P07) and
between 2008 July 13 and August 20 ({\it outer sample}). In both cases, FLAMES
has been used in GIRAFFE mode, using the high-resolution (R $\sim$ 22,500)
grating HR13 (6120-6395 \AA) and reaching a signal-to-noise ratio of $50-300~
pixel^{-1}$, depending on the star magnitude. The {\it inner sample} has been
already used by P07 to study the rotation pattern of the stellar populations of
$\omega$ Cen. The data reduction and calibration procedure is very similar to
that adopted here for the {\em outer sample} (see below), and is described in
detail in P07.

\begin{figure}
 \includegraphics[width=8.7cm]{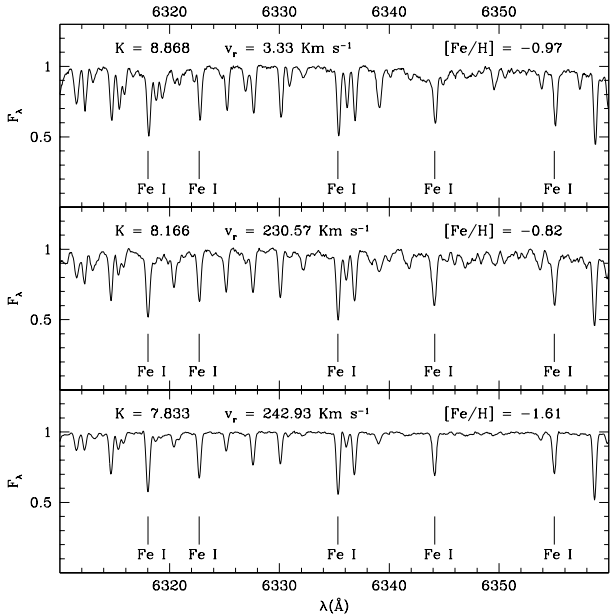}
\caption{Spectra of the field star 2MASS13331099-4749262 (top panel), and of the 
$\omega$ Cen stars 2MASS13262366-4742424 (middle panel) and 2MASS13283381-4732055 (bottom panel). 
A sample of iron lines used in the analysis are also indicated.}
\label{spec}
\end{figure}

The {\it outer sample} data have been reduced with the GIRAFFE BLDRS 
(Base-Line Data Reduction Software) 4 which includes cosmic-ray removal, bias 
subtraction, flat-field correction, wavelength calibration, and pixel 
resampling. Sixteen fibers have been dedicated to sky observations in each
exposure. An average sky spectrum has been obtained and subtracted from the
object spectra by taking into account the different fiber transmission. The
spectra have been then continuum-normalized and corrected for telluric 
absorption bands with IRAF. The spectra of three program stars are shown in
Figure \ref{spec} to illustrate the quality of our data.

Radial velocities have been derived through Fourier cross-correlation, using
the {\it fxcor} task in the radial velocity IRAF package. The spectrum of
each object has been correlated with a high signal-to-noise template spectrum
of the Geneva radial velocity standard star HD42807, retrived from the ELODIE archive (Moultaka et al.
2004). All spectra have been corrected for heliocentric velocity\footnote{The derived radial 
velocities of the {\it outer sample} stars are available in electronic form at the
CDS (http://cdsweb.u-strasbg.fr/).}. The average
error on the derived radial velocity is $\sigma_{v}\sim0.5~Km~s^{-1}$. Nine
stars are in common between the {\it inner} and the {\it outer} samples: 
the average difference in radial velocity is $\Delta~v_{r}=-0.09\pm0.16$, fully
consistent with no systematic shift between the two samples. 
Star by star comparisons with other datasets are also very satisfying.
The average radial velocity difference of the 278 stars in common with 
Reijns et al. (2006) is $\Delta~v_{r}=1.48\pm1.76$, 
while for the 36 stars in common with Johnson et al. (2008) is 
$\Delta~v_{r}=1.62\pm1.47$, i.e., consistent with null difference within the
uncertainties (Rejins et al. velocities were derived from spectra of resolution 
$8500<R<17000$; the resolution of Johnsons et al. spectra was R=13000;
in both cases the typical uncertainty is $\sim 1.5~Km~s^{-1}$).

In addition, metallicities have been also derived for the {\it outer sample}
stars. A set of 15 iron lines has been selected from the database of Kurucz \&
Bell (1995) in the spectral range covered by our spectra. These lines have
been identified, whenever possible, on each spectrum and have been fitted with
Gaussian functions. The integral of the difference between the continuum and
the line profile provided the equivalent width (EW) of each line. The
abundance analysis has been performed using the latest version of the Kurucz
(1979) model atmospheres and the MOOG line analysis code (Sneden 1973) to
compute LTE abundances from individual EWs. The abundance has been derived
from the model that best reproduced the observed EWs, for assumed values of
temperature, gravity, and microturbulence velocity. Temperatures have been
derived from J-K colors adopting the color-temperature transformations by
Montegriffo et al. (1998), a reddening $E(B-V)=0.11$ (Lub 2002) and the
extinction coefficients by Savage \& Mathis (1979). The gravity parameter
$log~g$ has been then computed assuming a distance modulus to $\omega$ Cen of
$(m-M)_{0}=13.70$ (Bellazzini et al. 2004; Del Principe et al. 2006). The
microturbulence velocity has been initially set as 2 $Km~s^{-1}$ and then
adjusted within a range of about 1 $Km~s^{-1}$ by minimizing the trend in the 
deduced abundances with EWs for Fe I lines. The average error on the derived
metallicities is $\sigma_{[Fe/H]}\sim0.2~dex$. The comparison with the
metallicity determinations by Johnson et al. (2008) for the 36 stars in common
indicates a difference of $\Delta [Fe/H]_{our-J}=-0.21\pm0.25$, indicating an
acceptable agreement for the purposes of the present study, that is focused on
the kinematics in the outer regions of $\omega$ Cen. Here we derived
metallicities of our stars to have an independent sanity check of the selection
criteria adopted in \S 3 to identify cluster members. A full analysis of the
abundance pattern of the selected stars will be the subject of a more detailed
dedicated analysis, that will be presented in a forthcoming paper.

\section{Radial velocity and Metallicity distribution}

\begin{figure}
 \includegraphics[width=8.7cm]{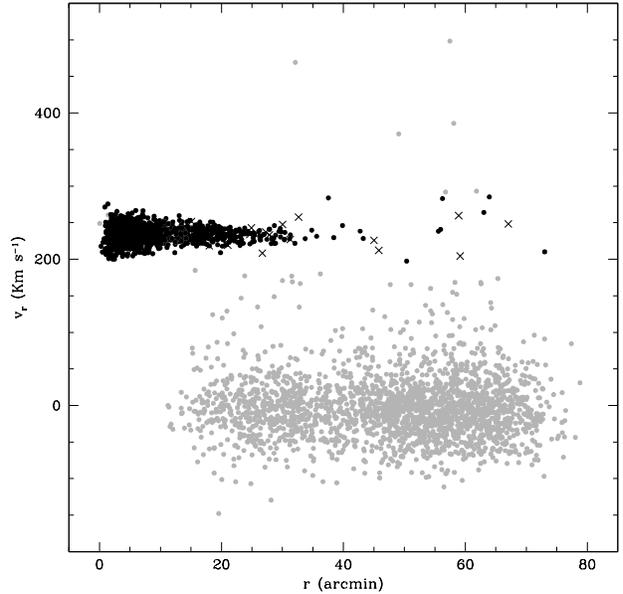}
\caption{The radial velocities of the entire sample are plotted as a function of
the distance from the cluster center. Grey points represent object rejected by
the radial velocity selection criterion ({\it i}) defined in Sect. \ref{sel}. Crosses
represent object rejected by the other selection criteria.}
\label{vel}
\end{figure}

In Figure \ref{vel} the radial velocities of the entire sample are plotted as a
function of the distance from the cluster center. To calculate distances 
we deprojected right ascension and declination into X and Y 
coordinates with the following relations, 
suited for extended objects that are not close to the celestial equator 
(from van de Ven et al. 2006): 

$$X=-r_{0}~cos~\delta_{0}sin(RA-RA_{0})$$
$$Y=r_{0}[sin~\delta~cos~\delta_{0}-cos~\delta~sin~\delta_{0}~cos(RA-RA_{0})]$$

where $RA_{0}$ and $\delta_{0}$ are the coordinate of the cluster center and 
$r_{0}=10800/\pi$ is the scale factor to have X and Y in arcmin. At the adopted
distance, 1 arcmin corresponds to 1.6 pc. As can be noted, the clump of
$\omega$ Cen stars is clearly visible at large positive velocities
($v_r>190~Km~s^{-1}$) while field stars are broadly distributed around the
average velocity $<v_r>=-8.0\pm 30.2~Km~s^{-1}$, in good agreement with the
predictions of the Galactic model by Robin et al. (2003; hereafter R03) for
thin disc and thick disc stars in this direction, and very different from the
typical velocity of $\omega$ Cen ($<v_r>=233.2\pm0.4~Km~s^{-1}$).

\begin{figure}
 \includegraphics[width=8.7cm]{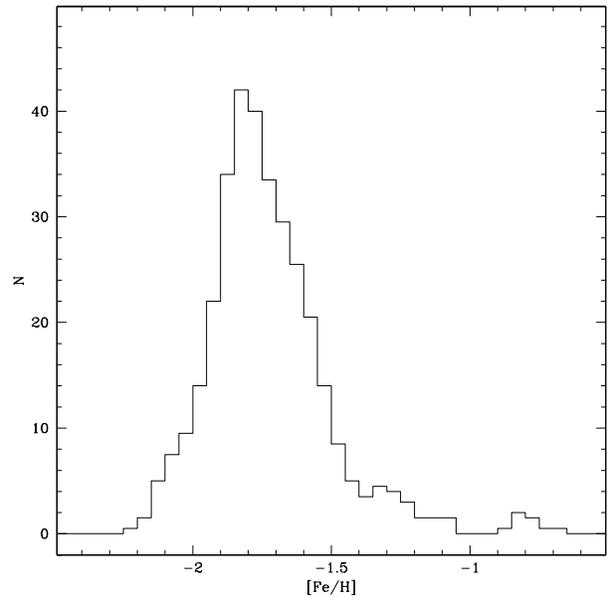}
\caption{Metallicity distribution of the 338 stars belonging to the {\it outer
sample} with radial velocity $190<v_r<290~Km~s^{-1}$. The distribution is
obtained using a running box of 0.1 dex width and a step size of 0.05 dex.}
\label{met}
\end{figure}

Figure \ref{met} shows the metallicity distribution obtained for the 338 stars
belonging to the {\it outer sample} and with radial velocity 
$190<v_r<290~Km~s^{-1}$. 
A sharp and asymmetric peak at $[Fe/H]\sim-1.8$ can be seen, with a long tail 
extending toward higher metallicity ($[Fe/H] > -1.4$). Secondary peaks are also
visible at metallicity $[Fe/H]\sim-1.3$ and $[Fe/H]\sim-0.85$, 
in agreement with the previous spectroscopic determinations by Norris et al.
(1996), Suntzeff \& Kraft (1996) and Johnson et al. (2008).

\section{Velocity dispersion}
\label{veldisp}

\subsection{Selection criteria}
\label{sel}

In order to construct the velocity dispersion profile of $\omega$ Cen we need
to distinguish the bona-fide cluster stars from the Galactic field
contamination. In fact, although the field population and $\omega$ Cen stars
have very different bulk velocities, a small (but non neglibible) 
fraction of stars belonging to the Galactic halo could lie in the same
velocity range of genuine cluster stars. This is particularly important in the
external region where the cluster density drops quickly with radius, while the
density of the field population remains constant (see \S \ref{out}).

\begin{figure}
 \includegraphics[width=8.7cm]{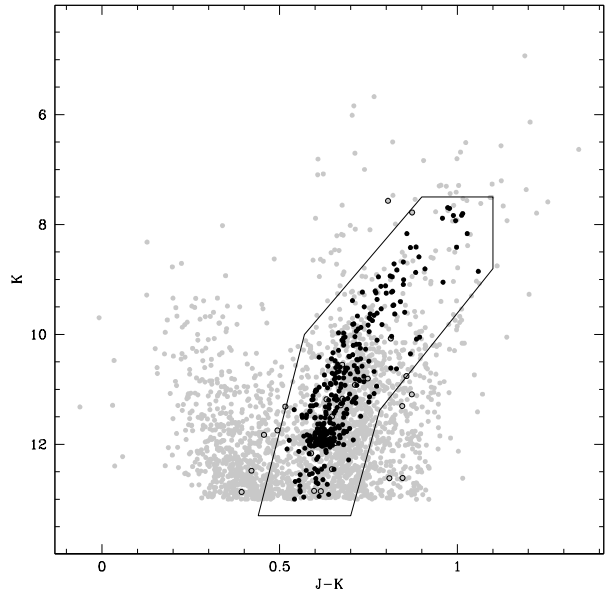}
\caption{$K~vs.~(J-K)$ CMD of $\omega$ Cen. Grey points represent the stars
rejected by the radial velocity selection criterion ({\it i})defined in Sect. \ref{sel}. 
Open points represent stars rejected by the other selection criteria. 
The adopted selection box is also shown. The clump of member stars around 
$K\simeq 11.7$ is the RGB bump of the cluster (see Sollima et al. 2004).}
\label{cmd}
\end{figure}

To this aim, we excluded from the {\it inner sample} all the stars with radial 
velocity $v_r<190~Km~s^{-1}$ (see P07) and adopted three 
different selection criteria for the {\it outer sample}:
\begin{itemize}
\item{}{\it i) Radial velocity:} The bulk radial velocity of $\omega$ Cen has been
estimated to be $<v_r>=233.2\pm0.4~Km~s^{-1}$ with a maximum velocity
dispersion at the center of $\sigma_{v}\simeq20~Km~s^{-1}$ (Meylan et al.
1995). Therefore, a first selection has been made by taking as possible members
only stars with
$190<v_r<290~Km~s^{-1}$\footnote{The choice of an asymmetric range has been
performed to minimize the field contamination in the region where the bulk of the field stars
are located.}. Note that this selection is safe since the
velocity dispersion is expected to decrease significantly at large distances
from the center. Therefore this criterion is not expected to introduce any bias
in the selected sample.
\item{}{\it ii) Position in the CMD:} Figure \ref{cmd} shows the $K~vs~(J-K)$ CMD of
the {\it outer sample} stars. We considered as cluster members only the stars 
lying within the selection box enclosing the cluster RGB that we have originally
adopted to chose high-priority targets, that is shown in Fig.~\ref{cmd}.
\item{}{\it iii) Proper motions:} We identified 245 stars in common with the proper
motion catalog by van Leeuwen et al. (2000). These authors provide for each star
a membership probability calculated according to its position in the proper
motions plane. On the basis of this additional 
information, we excluded all the 20 stars with a membership probability smaller  than 50\%.
\end{itemize} 

A total of 946 stars passed the above selection criteria, 628 belonging to the
{\it inner sample} and 318 belonging to the {\it outer sample}.
In Figure \ref{vel} the accepted and rejected stars are clearly indicated. 
The adopted approach is rather conservative as our aim is to obtain the most
reliable sample of cluster members to study the kinematics in the outer regions
where the velocity dispersion is low and can be biased by a limited number of
spurious sources. On the other hand, while the second and, in particular, the
third criterion listed above may exclude from the sample some genuine member,
this cannot bias our results as these selections are independent of the radial
velocity of the stars. Nevertheless, is interesting to note that using the R03
Galactic model as decribed in \S \ref{out}, below, we expect
a very low degree of contamination. In particular, the predicted number of
Galactic stars with $190<v_r<290~Km~s^{-1}$  in our sample is
$2.2\pm 0.5$  for $r\le 15\arcmin$, $1.4\pm 0.3$ for $15\arcmin <r\le
28\arcmin$, and
$1.4\pm 0.3$ in the range $28\arcmin <r\le 44\arcmin$ (see \S \ref{out} for
a discussion of the $r>44\arcmin$ region).

Another potential selection criterion is based on the metallicity. Note,
however, that only Galactic halo stars are expected to contaminate the sample
at velocities $190~Km~s^{-1}<v_r<290~Km~s^{-1}$. This Galactic component have a
metallicity distribution ranging between $-2.2<[Fe/H]<-0.8$ with a peak at
$[Fe/H]=-1.6$ (Ivezic et al. 2008) Moreover, the contaminating halo stars lie
in the foreground and background of $\omega$ Cen. Their different distances
translate into a wrong determination of their surface gravities and,
consequently, spread out their original metallicity distribution. As a
consequence, the resulting metallicity distribution of the halo population
largely overlaps the metallicity distribution of the $\omega$ Cen stars.
Therefore, this selection criterion is poorly efficient in discriminating the
cluster membership. However, we considered also a sample retaining only the
stars whose metallicity is close to the main peak of the metallicity
distribution shown in Figure \ref{met} ($-1.95<[Fe/H]<-1.55$). In the
following paragraph we calculated the velocity dispersion profile of $\omega$
Cen with and without adopting this last criterion, as a further check of our
results. 

\subsection{Method}
\label{method}

Once selected the bona-fide cluster member stars, as a first step we corrected
for perspective rotation. In fact, since $\omega$ Cen has a large extension on the 
plane of the sky (with a diameter about twice that of the full moon), its systemic 
motion produces a non-negligible amount of apparent rotation (Feast 1961).
We corrected for this effect using the following correction (see van de Ven et
al. 2006):

$$ \Delta v_r=0.001379~(X\mu_{X}+Y\mu_{Y})D~Km~s^{-1}$$

to be subtracted to the observed $v_r$.
$\mu_{X}=3.88$ and $\mu_{Y}=-4.44$ are the systemic proper motions in
units of $mas~yr^{-1}$ (van Leeuwen 2000) and $D=5.5$ is the distance to
$\omega$ Cen in kpc (Bellazzini et al. 2004; Del Principe et al. 2006). 

\begin{figure}
 \includegraphics[width=8.7cm]{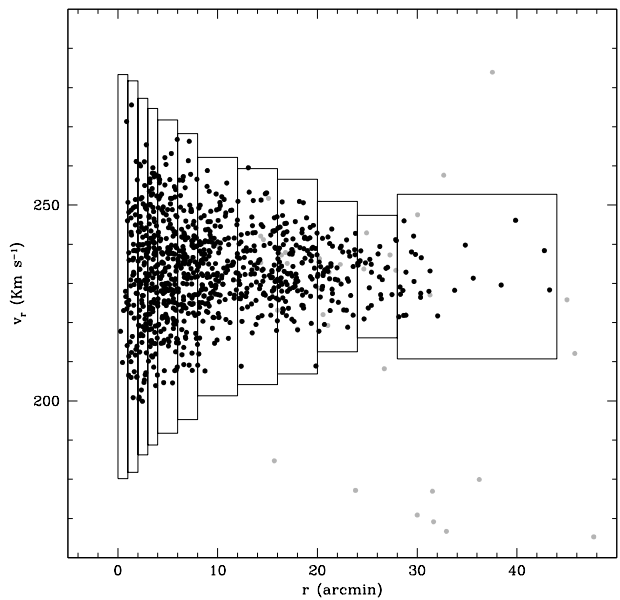}
\caption{Radial velocity distribution as a function of the distance from the
cluster center for stars with $190<v_r<290~Km~s^{-1}$. The adopted
bins of variable size are indicated together with the local $3\sigma$ range.
Grey points represent the object rejected by the selection criteria defined in
Sect. \ref{sel}}
\label{vsho}
\end{figure}

As a second step, we divided the sampled area in a set of concentric annuli at
various distances from the cluster center, searching for the best compromise to
have well populated and compact radial bins (except for the innermost
$r<1\arcmin$ bin that is not relevant in this context).
Within each annulus we calculated the value of $\sigma_{v}$ that maximize
the log-likelihood 

$$ L(\sigma_{v})=\sum_{i=1}^{N}log
\int_{-\infty}^{+\infty}
\frac{e^{-\frac{(v_{i}-<v>)^2}{2\sigma_{v}^{2}}}e^{-\frac{(v_{i}-<v>)^2}{2\sigma_{i}^{2}}}}{2\pi~\sigma_{v}~\sigma_{i}}dv
$$
where $v_{i}$ and $\sigma_{i}$ are the radial velocity and its corresponding
error for the N stars contained in the annulus and $<v>$ the average cluster
velocity (Walker et al. 2006). 
As a further selection, in each annulus we excluded the outliers stars whose 
radial velocity differs more than 3 times the {\em local} $\sigma_{v}$ 
from the cluster bulk velocity (see Figure \ref{vsho}; Bellazzini et al. 2008). 

\begin{figure}
 \includegraphics[width=8.7cm]{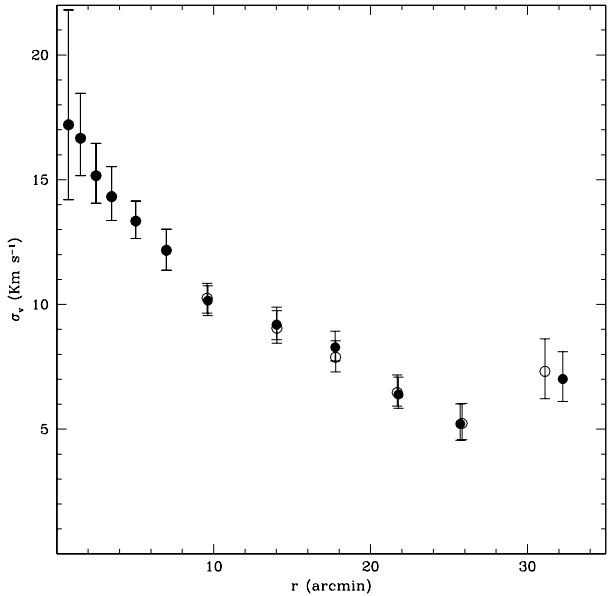}
\caption{Velocity dispersion profile of $\omega$ Cen. Open and filled points 
refers to the sample selected and not selected in metallicity, respectively.}
\label{sig}
\end{figure}

The derived velocity dispersion profile is listed in Table 1 and shown in
Figure \ref{sig} for the case of the sample selected with (filled points) and
without (open points) the metallicity selection criterion. As can be noted, no
significant differences are evident between the two profiles. For this reason,
in the following we refer always to the sample not selected in metallicity. The
obtained profile shows a central velocity dispersion of
$\sigma_{v}=17.2^{+4.6}_{-3.0}Km~s^{-1}$ in agreement with the previous 
estimates by Meylan et al. (1995), Mayor et al. (1997) Norris et al. (1997) and
van de Ven et al.(2006). The measured velocity dispersion appears to decrease
monotonically with distance out to $r\simeq 26\arcmin$ from the cluster center,
where it reaches a minimum of $\sigma_{v}=5.2^{+0.80}_{-0.65}Km~s^{-1}$. A small, non
statistically significant, increase of the velocity dispersion is noticeable
in the last bin, at a distance of $\sim 32\arcmin$. 

\begin{figure}
 \includegraphics[width=8.7cm]{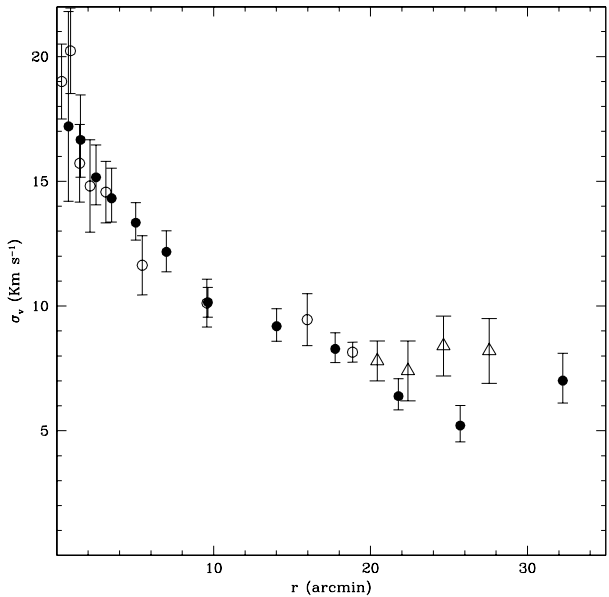}
\caption{Comparison between the velocity dispersion profile presented in this
paper (filled points) with those by van de Ven et al. (2006; open points) and by
Scarpa et al. (2003; open triangles).}
\label{comp}
\end{figure}

In Figure \ref{comp} the obtained velocity dispersion profile is compared with
the profiles obtained by van de Ven et al. (2006) and Scarpa et al. (2003).
Note that our profile agree very well with that obtained by van de Ven et al.
(2006) in the overlapping region $r\le 20\arcmin$. At distances
$r>24'$ the behaviour of our profile differs from that obtained by Scarpa et
al. (2003). In particular, while the profile by Scarpa et al. (2003) flattens
around a value of $\sigma_{v}\sim 8~Km~s^{-1}$, our profile continues to
decrease. The statistical disagreement between the two profiles is $\sim
2\sigma$, hence it cannot be considered as statistically significant.
The number of target stars in range $20\arcmin \la r\la 28\arcmin$ and the
accuracy of the radial velocity estimates of the present study is similar to
Scarpa et al. (2003). However, Scarpa et al. (2003) do not report (a) if their
velocities have been corrected for perspective rotation (\S 4.2) and (b) the
details of the adopted selection criteria; in particular it is not said if the
selection in $v_r$ takes into account the {\em local} value of the velocity
dispersion at the considered radius.
In any case, our results indicate that the flattening of the velocity
dispersion curve shown by Scarpa et al. (2003) may not reflect a real kinematic
feature of the cluster but a chance fluctuation, instead.
The (doubtful) increase of the dispersion in the $<r>\simeq 32\arcmin$ bin
observed here, if real, would be more compatible with the onset of tidal heating
that with the effects of MOND or DM (see \S \ref{nbsec}; see also Mu\~noz et al. 2008, and references
therein). Note that this last bin is located outside the radial range covered by Scarpa
et al.'s data.

\begin{table}
 \centering
 \begin{minipage}{140mm}
 \caption{Velocity dispersion profile of $\omega$ Cen.}
 \begin{tabular}{@{}ccccr@{}}
 \hline
  r   &  $<r>$  & N & $\sigma_{v}$ & $\Delta~\sigma_{v}$ \\
 arcmin &  arcmin  &  & $Km s^{-1}$   & $Km s^{-1}$ \\
 \hline
 0 - 1 &  0.75   & 10 & 17.20 & $^{+4.60}_{-3.00}$ \\
 1 - 2 &  1.51   & 50 & 16.66 & $^{+1.80}_{-1.50}$ \\
 2 - 3 &  2.50   & 80 & 15.16 & $^{+1.30}_{-1.10}$ \\
 3 - 4 &  3.49   & 91 & 14.32 & $^{+1.20}_{-0.95}$ \\
 4 - 6 &  5.04   & 151 & 13.34 & $^{+0.80}_{-0.70}$ \\
 6 - 8 &  6.99   & 112 & 12.17 & $^{+0.85}_{-0.80}$ \\
 8 - 12 & 9.63   & 150 & 10.15 & $^{+0.60}_{-0.60}$ \\
 12 - 16 & 14.01  & 98 & 9.19 & $^{+0.70}_{-0.60}$ \\
 16 - 20 & 17.75  & 99 & 8.28 & $^{+0.65}_{-0.55}$ \\
 20 - 24 & 21.79  & 48 & 6.39 & $^{+0.70}_{-0.55}$ \\
 24 - 28 & 25.71  & 24 & 5.21 & $^{+0.80}_{-0.65}$ \\
 28 - 44 & 32.25  & 26 & 7.01 & $^{+1.10}_{-0.90}$ \\
 \hline
\end{tabular}
\end{minipage}
\end{table}

\section{Comparison with Theoretical Models}

To check if canonical dynamical models are able to reproduce the observed
kinematics of $\omega$ Cen, in the following sections we compare our data with 
a set of analytical models (Wilson 1975). N-body simulations have also been
performed to study the effects of the tidal field of the Milky Way on different
models of the cluster.

\subsection{Analytical models}
\label{wils}

We attempt to reproduce our observations using Wilson (1975) models.
These models have the advantage of taking into account the cluster rotation, 
predicting simultaneously the surface brightness, the velocity dispersion, 
the rotation, and the ellipticity profiles.
These models adopt the distribution function

$$ f(E,J)=
\left\{
\begin{array}{lr}
(e^{-E}-1+E)e^{\beta J - \frac{1}{2}\zeta^{2}J^{2}} & \mbox{if}~E<0\\
0 & \mbox{if}~E \geq 0
\end{array}
\right.
$$

Any choice of the a-dimensional central potential $U_{0}$, and of the two free 
parameters of the above equation ($\beta$ and $\zeta$), yelds 
a different rotating model in virial equilibrium. The adoption of a
core radius $r_{c}$ and of the inclination angle respect to the plane of 
the sky $i$ provide an actual realization of the model that can be compared with
observations.
We explored the parameter space to find out one set reproducing 
simultaneously {\it i)} the surface brightness profile, {\it ii)} the velocity
dispersion profile, {\it iii)} the rotation curve and {\it iv)} the ellipticity
profile, giving lower weight to this last item.
We adopt the velocity dispersion profile derived in Sect. \ref{method}, 
the surface brightness profile by Trager, King \& Djorgovski (1995)
and the ellipticity profile by Geyer, Nelles \& Hopp (1983). Since the angle of
rotation in the plane of the sky of $\omega$ Cen is $\phi\sim0^{\circ}$, 
the rotation curve corresponds to the $X-v_r$ diagram.

\begin{figure*}
 \includegraphics[width=12cm]{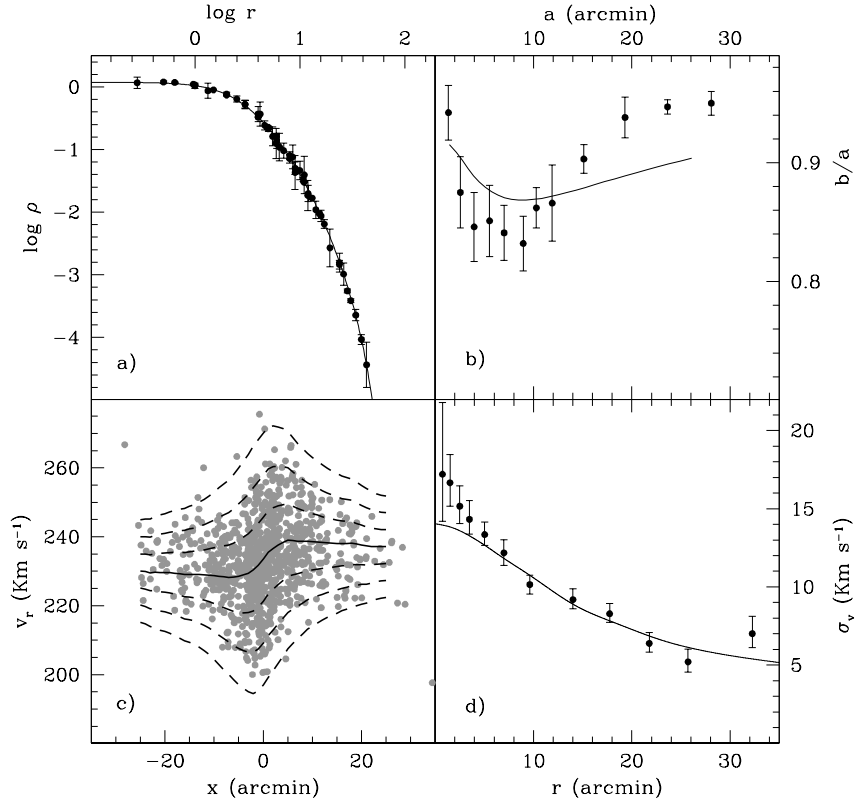}
\caption{Comparison between the best-fit Wilson (1975) model (solid lines) and 
a) the projected surface brightness (from Trager et al. 1995) b) the ellipticity
profile (from Geyer et al. 1983) c) the rotation curve and d) the velocity 
dispersion profile presented in this paper. In panel c) the $1\sigma$,$2\sigma$ 
and $3\sigma$ contours predicted by the model are also shown as dashed lines.}
\label{fit}
\end{figure*}

The best-fit model has been found by fixing the inclination angle $i=48^{\circ}$
(van de Ven et al. 2006) and assuming $U_{0}=-6.3$, $\beta=0.8$, $\zeta=0.4$ and
$r_{c}=2.8'$. The ratio between the rotational kinetic energy and the absolute
value of the potential energy of this model is $T/|W|=0.09$, indicating that it
is stable against non-axisymmetric perturbations (Ostriker \& Peebles 1973).
The fit of the observational data with this model is shown in
Figure \ref{fit}. As can be noted, the agreement is good in all the four
diagrams. In particular, the model well reproduces all the observed kinematic
properties of the cluster (rotation and dispersion) over the entire 
radial range covered by our data\footnote{The marginal disagreement between the
predicted and observed velocity dispersion profile in the innermost
region could be induced by many second-order effects in modelling the shape of
the velocity distribution in the internal cluster region (e.g. non-rotational
anisotropy) and/or by the possible presence of an intermediate-mass black hole object in the center of
$\omega$ Cen (as proposed by Noyola, Gebhardt \& Bergmann 2008)}.
This means that {\it the velocity
dispersion and rotation profiles of $\omega$ Cen are fully consistent 
with an equilibrium model in which the mass distribution follows the light,
in the framework of classical Newtonian gravitation.}
This is in good agreement with the results by McLaughlin \& Meylan (2003), that
were limited to the velocity dispersion profile.

\subsection{N-body simulations}
\label{nbsec}

As described in Sect. \ref{intro}, there is no strong evidence for the presence
of a significant population of tidally stripped stars around $\omega$ Cen. 
The most thorough study in this sense is the one by Da Costa \& Coleman (2008) 
that, from their extensive radial velocity survey, 
concluded that less than 0.7\% of
the total cluster mass is comprised between 1 and 2 tidal radii. This result imply 
that the stripping process must have been largely completed at early epochs.
Therefore, if $\omega$ Cen is indeed the remnant of a larger system destroyed 
long time ago, it must have
lost a large part of its stars in a orbit with a larger apogalacticon with respect
to the present day, because of orbital decay due to dynamical
friction (Zhao 2002; Bekki \& Freeman 2003; see also Tsuchiya et al. 2004; 
Mizutani et al. 2003 and  Ideta \& Makino 2004). 
This implies that the present-day system should be approximately in
equilibrium with the tidal field of the Milky Way and lie on a stable
orbit, as its mass is now sufficently low to make the effect of dinamical friction
negligible. 

We adopt this view as a working hypothesis for a simple N-body simulation aimed
at studying the spatial distribution and the kinematical properties of the stars
eventually stripped from the cluster by the tidal field of the Galaxy while the
system is in the present-day stable orbit (see Ideta \& Makino 2004). 
The very good knowledge of the
distance and 3-D motion of $\omega$ Cen (van Leeuwen 2000) gives a high
predictive power, under the limitations of the assumed hypothesis.
For more detailed and extensive studies see Bekki \& Freeman (2003), 
Tsuchiya et al. (2004), Ideta \& Makino (2004), and references therein. 

The simulation was performed with {\tt falcON}, a fast and momentum-conserving
tree-code (Dehnen 2000, 2002), within the NEMO environment (Tauben 1995). 
Gravity was softened with the kernel `P$_2$' of Dehnen (2000), with a softening
length of 2 pc. The model cluster was let to evolve within the  three-component
(bulge+disc+halo) static Galactic potential 2b of Dehnen \& Binney (1998),
which has $v_{rot}(R_{\sun})=231~Km~s^{-1}$ and a total mass within 100 kpc of
$\simeq 1 \times 10^{12} M_{\sun}$. $\omega$~Cen was represented by 50000
particles distributed according to an equilibrium King (1966) model having
$W_0=6.0$, $r_t=100$ pc and $M=5\times 10^6 ~M_{\sun}$. This model has
concentration ($c$) and tidal radius ($r_t$) values similar to the real
cluster; the central velocity dispersion is also very similar, $\sigma\simeq
17.0~Km~s^{-1}$, but it should be recalled that the initial velocity
distribution is completely isotropic, lacking any intrinsic rotation. {\tt
falcON} is best suited to follow the evolution of non-collisional systems.
While the effects of two-body encounters have been shown to be negligible in
$\omega$~Cen (Ferraro et al. 2006; Sollima et al. 2007), this may not be the
case in the innermost region of our model.  To minimize undesired effects due
to two-body encounters in the densest part of the model we have adopted a
softening length approximately as large as the cluster  core radius.  For this
reason the evolution in the innermost few $r_c$ of the model should not be
considered as a good representation of the evolution of the real cluster. On
the other hand our model is fully adequate to study the reaction of the outer
regions to the tidal strain of the Galactic potential, i.e. the morphology and
the kinematics of possible tidal tails, that is what we are interested into.
Note also that the 2-body relaxation time at the half-mass radius of our model (Binney \& Tremaine
1998, their Eq. 4-9) is a factor of $\simeq 4$ larger than the duration of our
simulation, hence, in any case, the effects of particle encounters on the
global properties of the model should be negligible.

\begin{figure}
 \includegraphics[width=8cm]{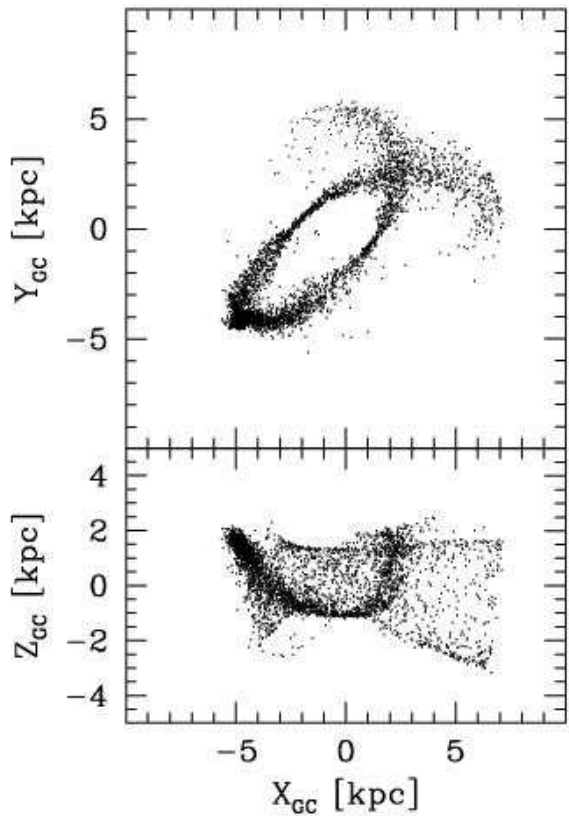}
\caption{Position of the particles at the end of the N-body simulation A (present
epoch) in galactocentric coordinates. The X-Y plane coincides with the Galactic
Plane, the X-Z plane is perpendicular to it and passes through the Sun. Note the
excellent agreement with the much finer simulation by Ideta \& Makino (2004;
the upper right panel of their Fig.~2).}
\label{vedox}
\end{figure}

First of all we integrated backward in time, within the adopted potential, the
orbit of a test particle whose initial position and 3-D velocity coincide with
those of $\omega$~Cen at the present epoch (position from Bellazzini et al.
2004 and velocity from van Leeuwen et al. 2000). The orbit has a period of
$\simeq 80$ Myr, pericentric and apocentric distances of 1.5 kpc and 6.5 kpc,
respectively, very similar to the orbits derived by Mizutani et al. (2003) and
Ideta \& Makino (2004).  We launched our N=50000 particle model from the
apocenter at the epoch t=0.972 Gyr in the past. With this assumption the
cluster evolves within the adopted potential for more than 10 orbits from that
time to the present epoch (t=0). At the end of the simulation the model is
within 40 pc of the current position of $\omega$~Cen, and the 3-D velocity is
within 5\% of the real cluster velocity. The distribution of the particles at
the end of the simulation (see Fig.~\ref{vedox}) is in good agreement with the
results by Ideta \& Makino (2004) and Mizutani et al. (2003). The remnant have
developed two tidal tails extending over the whole orbital path. However, the
tidal radius of the bound remnant is only 10\% smaller than the initial model
and there is a mere 8\% of the particles that lie outside that radius after 10
perigalactic passages. This confirms that $\omega$~Cen is currently nearly at
equilibrium with the tidal field of the Milky Way  and that any tidal tail
originating from the present day cluster should be very weak and hard to
detect. In particular, it is interesting to note that only 0.4\% of the
particles are found in the annulus between 1 $r_t$ and 2 $r_t$, in excellent
agreement with the results of Da Costa \& Coleman (2008). Rescaling to the
observed total luminosity of the cluster the average surface brightness in that
annulus due to tidally stripped stars is $\mu_V\sim 30.0 $ mag/arcsec$^2$.

\begin{figure*}
 \includegraphics[width=12cm]{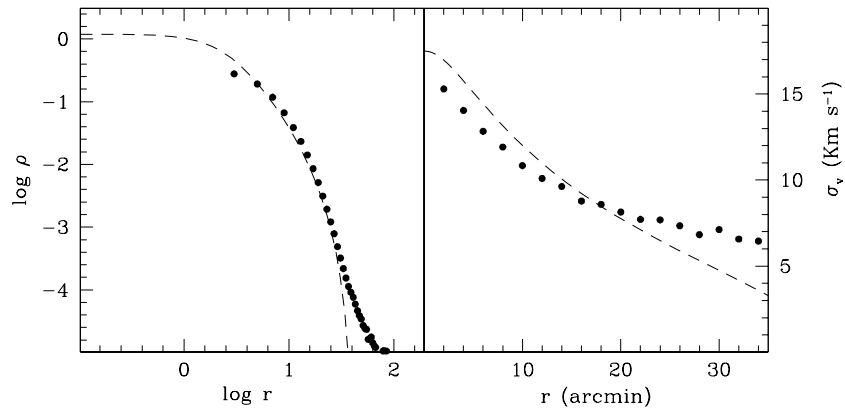}
\caption{Surface brightness (left panel) and velocity dispersion (right panel) 
profiles for the bound remnant at the end of the N-body simulation A (filled
circles). The dashed lines are from the King (1966) model that best fits the
final surface brightness profile.}
\label{nbody}
\end{figure*}

In Figure \ref{nbody} the final surface brightness and the velocity dispersion
profiles of the simulated satellite are shown. As can be noted, the surface
brightness profile deviates from the King profile at $r>40\arcmin$, following 
the power-law trend typical of tidally stripped stars (see Johnston, Sigurdsson
\& Hernquist 1999; Mu{\~n}oz et al. 2008, and references therein). This
deviation occurs at a very low surface density level (about 3.5 orders of
magnitude lower than the center), just outside the limit reached so far by any
photometric investigation. Note that neither King models nor Wilson's ones can
reproduce this external behaviour.

Also the velocity dispersion profile shows a
departure from the predicted King profile. However, in this case the drift
already occurs at $r\sim 25\arcmin$ yielding a flatter profile at larger
distances. In any case, the derived profile is compatible with the observed one
, within the uncertainties. Hence, the available observations appear still
compatible with the tidal stripping scenario described by our N-body model.
Fig.~\ref{nbody} suggests that the ``easiest'' way to look for the
observational signatures of the ongoing formation of tidal tails would be to
reliably extend the surface brightness profile using abundant and
easy-to-select tracers as Main Sequence stars, searching for the power-law
branch of the profile. 

\section{The outer region}
\label{out}

In Sect. \ref{veldisp} we presented the velocity dispersion profile of $\omega$
Cen calculated out to a distance of $\sim$32 arcmin, and comprising stars out to
$r=44\arcmin$ from the cluster center.
A more careful analysis is needed when considering the region at 
$r>44'$ i.e. close to the cluster tidal radius.
In this region, only seven stars passed the selection 
criteria defined in Sect. \ref{sel}. 
The resulting velocity dispersion of this sample is
$\sigma_{v}=36.0^{+12.5}_{-7.5}~Km~s^{-1}$, about a factor of 2 larger than the
central value, and a factor of 5 larger than the outermost point of the profile, at
$<r>\simeq 32$. It is worth noticing that although the adopted selection
criteria are efficient in eliminating the contaminating field stars in the inner
region of the system, their efficiency decrease at larger distances, as the surface
density of $\omega$ Cen stars plunges below the background level due to Galactic
field stars. 

To check the membership of these sample of stars we performed various tests.
First, we compared our sample with the predictions of the R03 Galactic model.
Although this model is only a statistical representation of the Galactic field 
population (see Drimmmel et al. 2005), it gives a good estimate of the order of magnitude of the field
contamination.
To minimize the effect due to small number statistics, we retrived (and merged
together) two simulated catalogue covering an area of 5 sqare degrees each, in the
direction of $\omega$ Cen. The $v_r$ distributions of simulated stars
approximately lying in the same magnitude and colour range of the observed
stars is well fitted by a sum of two Gaussian distributions, one for the main
peak of thin disc + thick disc Galactic stars at $-150<v_r<150~Km~s^{-1}$, the
other for the Halo population. Once fixed the density normalization of the two
components, the number of stars expected in our sample, in any given range of
$r$ and $v_r$ is  easily obtained by re-normalizing the Gaussian of the main
component to the number of observed stars having $-150<v_r<150~Km~s^{-1}$ in
that range. Adopting this procedure, the expected number of field stars with
$190<v_r<290~Km~s^{-1}$ at distances $r>44\arcmin$ turns out to be
$N_{field}=4.0\pm0.9$. Therefore, according to this test, about half of the
outermost stars are likely non-members, Galactic Halo stars.

Another possibility is to estimate the number of cluster stars expected at such
distance by means of the cluster density profile. Indeed, in a given radial bin (defined
between $r_{1}$ and $r_{2}$) the expected ratio between the number of cluster stars
($N_{mem}$) and field stars ($N_{field}$) can be written as

\begin{equation}
\frac{N_{mem}}{N_{field}}=\frac{\int^{r_{2}}_{r_{1}} 2 \pi r \rho_{mem}
dr}{\int^{r_{2}}_{r_{1}} 2 \pi r \rho_{field} dr}=C~\frac{\int^{r_{2}}_{r_{1}} r \rho_{mem} dr}{\int^{r_{2}}_{r_{1}} r~dr}
\label{eq1}
\end{equation}

where $\rho_{mem}$ and $\rho_{field}$ represent the projected surface density of
member stars and field stars, respectively. In the above equation the density of
field stars has been assumed to be constant over the field of view of the cluster.
We counted $N_{field}$ as the number of stars with 
$190~Km~s^{-1}<v_r<290~Km~s^{-1}$ in each radial bin and estimated the constant
$C$ by best-fitting the ratio $N_{mem}/N_{field}$ measured in the innermost bins
(at $r<44'$) with the prediction of the best-fit Wilson (1975) model described in
Sect. \ref{wils}. Then, we estimated the number of cluster members stars in the
most external bin by using eq. \ref{eq1} and $N_{field}$ measured in this bin. 
Following this procedure we estimated $N_{mem}=0.4$ at $r>44\arcmin$. However, as shown
in Sect. \ref{nbsec}, the interaction between $\omega$ Cen and the Milky Way would
produce a departure from the King profile in the external region of the cluster,
implying a higher surface density. Therefore, the number of cluster members stars
in the external region is expected to be larger than this estimate. We performed
the above procedure also assuming at $r>40'$ a power-law density profile with
index $\alpha=-3.8$ (i.e. the slope predicted by our N-body simulation). Also in
this case, the expected number of cluster members results $N_{mem}=1.5$. Thus,
according to this test, the majority of the high-velocity external stars should
be field stars.

It is interesting to note that none of the seven considered stars pass the
metallicity selection criterion defined in Sect. \ref{sel}. 
A Kolmogorov-Smirnov
test gives a probability smaller than 0.2\% that the metallicity of
these stars can be randomly extracted from the metallicity distribution of the
entire sample of cluster members stars shown in Figure \ref{met} .
Given all the result above we can safely conclude that most of (if not all) 
the seven stars passing our selection criteria at $r>44\arcmin$ are likely 
not member of the cluster.

\begin{figure}
 \includegraphics[width=8.7cm]{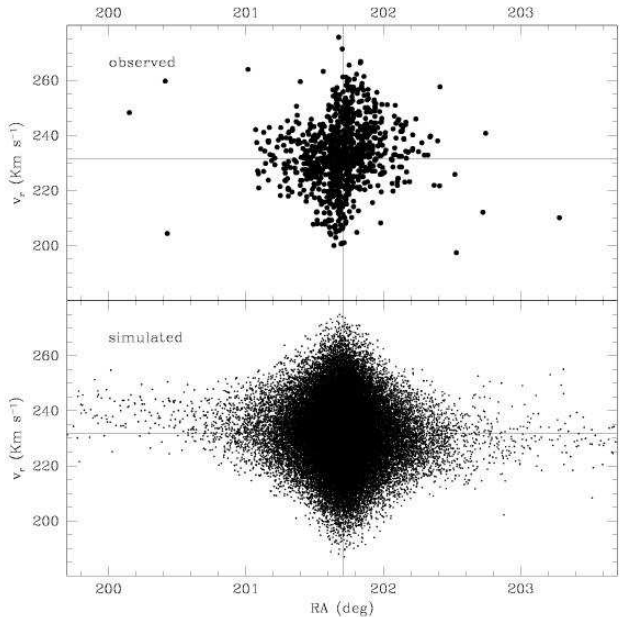}
\caption{The radial velocities of the stars with
$190<v_r<290~Km~s^{-1}$ is shown as a function of the right ascension for the
observed sample (top panel) and for the N-body simulation described in Sect. \ref{nbsec}
(bottom panel).}
\label{tidal}
\end{figure}

Having established that, we draw the attention of the reader on a curious
occurrence. In Figure \ref{tidal} the radial velocities of the stars with
$190<v_r<290~Km~s^{-1}$ is shown as a function of the right ascension for the
observed sample and for the particles of the N-body model described in Sect. 
\ref{nbsec}. Note that the most external stars are not evenly distributed in
the considered plane. Most of the $r>44\arcmin$ stars lying to the West of the
cluster have $v_r$ larger that the systemic cluster velocity, while the
opposite is true for $r>44\arcmin$ Eastern stars. This apparent trend of mean
$v_r$ with RA is in the same sense as that expected to be imprinted by the
Galactic tidal field on stripped stars, clearly visible in the N-body model
(lower panel of Fig.~\ref{tidal}). By means of Monte-Carlo extractions from the
synthetic R03 catalogue described above, we estimated that the chance
occurrence of such an asimmetry in Halo stars is lower than 5\%. If the
observed gradient would be revealed as real using a much larger sample of stars
in this range of  distances, this would make very difficult to interpret most
of these stars as belonging to the Galactic Halo and the hypothesis of a tidal
tail must be reconsidered. 
Another interesting issue emerges from this comparison: in the innermost
region of the cluster ($201.0\degr \la RA\la 202.3\degr$ in Fig\ref{tidal}),
the clear S-shaped trend due to the cluster rotation is clearly visible
In particular, in this case the Eastern branch of the rotation curve has larger
average $v_r$ with respect to the Western branch.
This trend is just in the opposite direction of what produced in the N-body
simulation by the Galactic tidal strain. This strongly suggest that the rotation
in the inner $30\arcmin$ of $\omega$ Cen is not induced the interaction 
with the Milky Way, as previously suggested by van de Ven et al. (2006), 
but it is an intrinsic property of the system, instead. 

\section{Summary and conclusions}

In this paper we present high accuracy radial velocities for a sample of more 
than 2500 stars in the direction of $\omega$ Cen. Adding the homogeneous sample by P07 and 
applying various un-biased selection criteria, we selected a sample of 946
bona-fide cluster's members displaced over a wide region up to a distance of
$\sim 80\arcmin$ from the cluster center.

We derived a very reliable velocity dispersion profile in the range 
$1.5\arcmin \le r\le 28\arcmin$. In this range the velocity dispersion appears to 
decrease monotonically from a central value of $\sigma_{v}\sim~17.2~Km~s^{-1}$ down to a
minimum value of $\sigma_{v}\sim~5.2~Km~s^{-1}$. A non-statistically significant
raise to $\sigma_{v}\simeq 7.0\pm 1.0~Km~s^{-1}$ is observed in an outer extended radial
bin ($28\arcmin\le r< 44\arcmin$), that must be considered with great caution
because of the uneven radial distribution of the 26 candidate member stars
included. The obtained profile is consistent with those obtained by previous
authors within $r\simeq 20\arcmin$ (Meylan et al. 1995; Mayor et al. 1997; van de
Ven et al. 2006) but differs from that derived by Scarpa et al. (2003) in the most
external region of the cluster. In particular, at odds with the Scarpa et al.'s
analysis, we did not find any sign of flattening of the velocity dispersion
profile in the region between $20\arcmin<r<28\arcmin$. 

The observed velocity dispersion profile of $\omega$ Cen, as well as its 
structural and kinematical properties is well reproduced by a Wilson (1975) 
equilibrium model in which mass follows light. Therefore, with the present data we
find no evidences of neither the presence of dark matter nor MOND effects, at
least in the inner $30\arcmin$ from the center.

The majority of the seven candidate members we found at $r>44\arcmin$ are
consistent with being iterlopers from the Galactic Halo. The only piece of evidence
that is in marginal contrast with this view is that they seem to display a velocity
gradient similar to that expected for star tidally stripped from $\omega$ Cen,
according to the prediction of a simple N-body model.
The comparison with the same N-body model suggest that the well-known 
rotation pattern observed in the inner $30\arcmin$ of $\omega$ Cen is not produced
by the Galactic tidal strain, but it should be intrinsical to the stellar system,
instead.

The present study, mainly focused on the kinematics in the range 
$20\arcmin \la r\la 30\arcmin$, as well as that performed by Da Costa and Coleman
(2008), focused on the search of extra-tidal stars 
($60\arcmin \la r\la 120\arcmin$), reached the limits that can be achieved using
RGB stars as tracers. Future studies aimed at assessing the shape of the velocity
dispersion profile in the range $30\arcmin \la r\la 60\arcmin$ and trying to
identify extra-tidal stars must rely on much more abundant Main Sequence stars.

\section*{Acknowledgments}

This research was supported by the Instituto de Astrofísica de Canarias. M.B.,
M.C., E.P., and R.S. acknowledge the financial fupport to this research by INAF
through the PRIN-2007 Grant CRA 1.06.10.04. We warmly thank Paolo Montegriffo
for assistance in the analysis. We also thank the anonymous referee for his
helpful comments and suggestions.

\bsp

\label{lastpage}

\end{document}